\documentclass[showpacs,preprintnumbers,two column]{revtex4-1}
\usepackage{amssymb}
\usepackage{amsmath}
\usepackage{graphicx}
\usepackage{dcolumn}
\usepackage{bm}

\begin{document}

\title{Analytical solutions and genuine multipartite entanglement of the
three-qubit Dicke model}
\date{\today}
\author{Yu-Yu Zhang$^{1,*}$, Xiang-You Chen$^{1}$, Shu He$^{2}$, Qing-Hu Chen%
$^{2,3,\dagger}$}
\address{
$^{1}$Department of Physics, Chongqing University, Chongqing
401331, P. R.  China\\
$^{2}$Department of Physics, Zhejiang University, Hangzhou 310027,
P. R. China\\
$^{3}$  Collaborative Innovation Center of Advanced Microstructures,
Nanjing 210093, China
}

\begin{abstract}
We present analytical solutions to three qubits and a single-mode cavity
coupling system beyond the rotating-wave approximation (RWA). The zeroth-order approximation, equivalent to the adiabatic approximation, works well for arbitrary coupling strength for small qubit frequency.
The first-order approximation, called the generalized
rotating-wave approximation (GRWA), produces an effective solvable
Hamiltonian with the same form as the ordinary RWA one and exhibits
substantial improvements of energy levels over the RWA even on resonance.
Based on these analytical eigen-solutions, we study both the bipartite
entanglement and genuine multipartite entanglement (GME).
The dynamics of these two  kinds of entanglements   using the GRWA are consistent with the
numerical exact ones. Interestingly, the well-known sudden death of entanglement
occurs in the bipartite entanglement dynamics but not in the GME dynamics.
\end{abstract}

\maketitle

\section{Introduction}

The interaction between qubits and a cavity is ubiquitous in several
branches of physics ranging from quantum optics~\cite{Scully}, to quantum information~\cite%
{Wallraff}  to condensed-matter physics~\cite{Leggett}. In early work on cavity quantum electrodynamics (QED),
the qubit-cavity coupling strength was much smaller than the cavity
transition frequency, the rotating-wave approximation (RWA) can be applied, and an analytical exact solution can be derived straightforwardly~\cite{jaynes}.
With recent advances in the
circuit QED using superconducting qubits, it is possible to engineer
systems for which the qubits are so far detuned from the cavity, or are
coupled to the cavity in a ultra-strong coupling regime where the coupling
strength is comparable to the cavity transition frequency, that the RWA is demonstrated  to
fail to describe the system correctly~\cite%
{Niemczyk,pfd,fedorov,devoret,you,li}. The counter-rotating-wave (CRW)
interactions in the qubit-cavity systems are therefore expected to play a crucial role.

Under the RWA, the ground state  is simply a direct product of the low state of the qubit and the vacuum cavity.
The CRW interactions lead to a squeezed vacuum state containing virtual photons~%
\cite{Ashhab,ciuti}. The analytical exact study in the full model is highly nontrivial. There have been numerous theoretical studies on
one- and two-qubit and cavity coupling systems, including the adiabatic approximation~\cite%
{Ashhab1,agarwal}, a Bargmann space technique~\cite{Braak,Zhong}, an
extended coherent-state method~\cite{chen,mao15}, and a generalized RWA (GRWA)~\cite{irish,zhang}. Recently there
have been interesting applications of the Dicke model~\cite{dicke} with three
qubits in the quantum information technology, such as the application of the Greenberger-Horne-Zeilinger states~\cite%
{GHZ}. And the circuit QED has entered the deep-strong-coupling
regime ~\cite{Yoshihara}, so it is experimentally possible to
realize the three-qubit Dicke model in circuit QED in the ultra-strong-
and deep-strong-coupling regime~\cite{sonalo}. We will present an
analytical solution to a three-qubit Dicke model. However, explicit
analytic solutions to the three- and more-qubit Dicke model have not
been extensively studied. Despite the fact that the exact solution to the
three-qubit  Dicke model has been given by a Bargmann space technique~\cite%
{Braak1} where a numerical search for the zeros of very complicated
transcendental  functions is needed, an efficient, easy-to-implement
theoretical treatment remains elusive. In this paper, we extend the
previous GRWA in  the one-qubit Rabi model by Irish~\cite{irish}  to
the three-qubit Dicke model. Including the CRW interactions, we
successfully derive a solvable Hamiltonian with the same form as the
ordinary RWA term. Therefore all eigenvalues and eigenstates can be
approximately solved and can be implemented with great ease by
experimentalists.

There is on going interest in the genuine multipartite
entanglement (GME) of the Dicke states for multiple qubits systems~\cite%
{novo,horodecki}.
Most of the existing studies of
entanglement focus on bipartite entanglement in the reduced state of two parties of a multipartite system~\cite%
{nielsen,roscilde,giamp,zurek}, which can be quantified through the von
Neumann entropy~\cite{wehrl,schu} and the concurrence characterizing qubit-qubit
entanglement~\cite{wang,vidal,zhang1}. However, bipartite entanglement can
only give a partial characterization. Multipartite entanglement is
known to be different from entanglement between all bipartitions~\cite%
{ali,Guhne,bodoky}. Recently the
bipartite entanglement decoherence has been studied in connection with a
phenomenon termed entanglement sudden death, indicating that the bipartite
entanglement can decay to zero abruptly during a finite period of time~\cite%
{yu1}. Whether this property occurs for the dynamics of GME remains
unexplored. So it is highly desirable to study both the bipartite
entanglement and the GME for the multipartite entanglement in the more than
two qubits system, where the three qubits and cavity coupling system can be
served as the most simple paradigm.

The paper is outlined as follows. In Sec.~II, we map the three-qubit Dicke
model with the CRW interactions into a solvable Hamiltonian by the zeroth- and
first-order approximation, giving an analytical expression of eigenvalues and
eigenstates. In Sec.III, we discuss dynamics of the GME for the multi-qubit
entanglement and the concurrence for the qubit-qubit entanglement by our
method. Finally, a brief summary is given in Sec.~IV.

\section{An analytical treatment to the three-qubit cavity system}

The Hamiltonian of the three-qubit Dicke model, which describes three
identical qubits coupled to a common harmonic cavity, is written as $\left(
\hbar =1\right) $
\begin{equation}
H=-\Delta J_{z}+\omega a^{\dagger }a+\frac{g}{2}(a^{\dagger
}+a)(J_{+}+J_{-}),  \label{ham1}
\end{equation}%
where $a$ and $a^{\dagger }$ are, respectively, the annihilation and
creation operators of the harmonic cavity with frequency $\omega $, $%
J_{i}\left( i=z,\pm \right) \ $\ is the angular momentum operator,
describing the three qubits of level-splitting $\Delta $ in terms of a
pseudospin of length $J=3/2$, and $g$ denotes the collective qubit-cavity
coupling strength.

In the RWA, the CRW terms $a^{\dagger }J_{+}$ and $aJ_{-}$ are neglected, and the Hamiltonian becomes
\begin{equation*}
H_{\mathtt{RWA}}=-\Delta J_{z}+\omega a^{\dagger }a+\frac{g}{2}(a^{\dagger
}J_{-}+aJ_{+}),
\end{equation*}
which is restricted to relatively weak-coupling strength $g\ll \omega $, and
to the qubit-cavity near resonance, $\Delta \approx \omega $. Now, the
interaction couples only $|-\frac{3}{2}\rangle |n+2\rangle $, $|-\frac{1}{2}%
\rangle |n+1\rangle $, $|\frac{1}{2}\rangle |n\rangle $, and $|\frac{3}{2}%
\rangle |n-1\rangle $ for each $n$, which are the eigenstates of the noninteracting Hamiltonian $-\Delta J_{z}+\omega a^{\dagger }a$. The whole Hilbert space can then be decomposed into the subspaces formed by these states which  can be diagonalized analytically. It is easy
to write the following tri-diagonal matrix form:
\begin{widetext}
\begin{equation}
H_{\texttt{RWA}}=\left(
\begin{array}{cccc}
\omega (n+2)+\frac{3\Delta }{2} & T_{n+1,n+2} & 0 & 0 \\
T_{n+1,n+2} & \omega (n+1)+\frac{\Delta }{2} & T_{n,n+1} & 0 \\
0 & T_{n,n+1} & \omega n-\frac{\Delta }{2} & T_{n-1,n} \\
0 & 0 & T_{n-1,n} & \omega (n-1)-\frac{3\Delta }{2}%
\end{array}%
\right) .  \label{RWA}
\end{equation}%
\end{widetext}
where
\begin{eqnarray*}
T_{n+1,n+2}&&=g\sqrt{3(n+2)}/4,T_{n,n+1}=g\sqrt{n+1}/4, \\
T_{n-1,n}&&=g\sqrt{3n}/4.
\end{eqnarray*}

If CRW terms $a^{\dagger }J_{+}$ and $aJ_{-}$ are included, the Hilbert
space cannot be decomposed into the finite dimensional spaces, because the
total excitation number $N=a^{\dagger }a+J_{z}+3/2$ is non-conserved and the
subspace for different index $n$ defined above is highly correlated. So
analytical solutions in this case should be highly non-trivial.

The Hamiltonian (\ref{ham1}) including the CRW terms with a rotation around the $%
y$ axis by an angle $\pi/2$ can be rewritten as
\begin{equation}
H=\Delta J_{x}+\omega a^{\dagger }a+g(a^{\dagger }+a)J_{z}.  \label{Ham}
\end{equation}%
Introducing a unitary transformation $U=\exp \left[ \frac{g}{\omega }%
J_{z}\left( a^{\dagger }-a\right) \right] $, one can obtain the transformed
Hamiltonian $H'_{SB}=H_{0}+H_{1}$, consisting of
\begin{eqnarray}  \label{H_11}
H_{0} &=&\omega a^{\dagger }a-\frac{g^{2}}{\omega }J_{z}^{2},  \label{H_0} \\
H_{1} &=&\Delta \left\{ J_{x}\cosh \left[ \frac{g}{\omega }\left( a^{\dagger
}-a\right) \right] +iJ_{y}\sinh \left[ \frac{g}{\omega }\left( a^{\dagger
}-a\right) \right] \right\}.\label{H_11}  \notag \\
\end{eqnarray}%
Then We can expand the even and odd functions $\cosh (y)$ and $\sinh (y)$, respectively, as
$\cosh \left[ \frac{g}{\omega }\left(
a^{\dagger }-a\right) \right] =G_{0}\left( a^{\dagger }a\right) +G_{1}\left(
a^{\dagger }a\right) \left( a^{\dagger }\right) ^{2}+a^{2}G_{1}\left(
a^{\dagger }a\right) +...$ and $\sinh \left[ \frac{g}{\omega }\left(
a^{\dagger }-a\right) \right] =F_{1}\left( a^{\dagger }a\right) a^{\dagger
}-aF_{1}\left( a^{\dagger }a\right) +F_{2}\left( a^{\dagger }a\right) \left(
a^{\dagger }\right) ^{3}-a^{3}F_{2}\left( a^{\dagger }a\right) +...$, where $%
G_{i}(a^{\dagger }a)(i=0,1,...)$ and $F_{j}(a^{\dagger }a)(j=1,2,...)$ are
coefficients that depend on the cavity number operator $\widehat{n}=a^{\dagger}a$
and the dimensionless parameter $g/\omega$. A different order of approximations can then be performed by
neglecting some terms in the expansions.

\textsl{zeroth-order approximation:} In the zeroth-order approximation, we
only keep the first term $G_{0}\left( a^{\dagger }a\right) $ in $\cosh %
\left[ \frac{g}{\omega }\left( a^{\dagger }-a\right) \right] $, and the Hamiltonian is approximated as
\begin{equation}
H^{^{0th}}=\omega a^{\dagger }a-\frac{g^{2}}{\omega }J_{z}^{2}+\Delta
J_{x}G_{0}\left( a^{\dagger }a\right) .
\end{equation}
In the basis of the oscillator state $|n\rangle$, the term $G_{0}\left( a^{\dagger }a\right) $ only has
non-vanishing diagonal element
\begin{equation}
G_{0}(n)=\langle n|\cosh \left[ \frac{g}{\omega }\left( a^{\dagger
}-a\right) \right] |n\rangle =e^{-\frac{g^{2}}{2\omega ^{2}}}L_{n}(\frac{%
g^{2}}{\omega ^{2}}),
\end{equation}
where Laguerre polynomials $L_{n}^{m-n}(x)=\sum_{i=0}^{\min
\{m,n\}}(-1)^{n-i}\frac{m!x^{n-i}}{(m-i)!(n-i)!i!}$.
Note that only the
oscillator number operator $\widehat{n}$ appears, so the
Hilbert space can be decomposed into different $n$ manifolds spanned by the spin
and cavity basis of $|-\frac{3}{2}\rangle |n\rangle $, $|-\frac{1}{2}\rangle
|n\rangle $, $|\frac{1}{2}\rangle |n\rangle $ and $|\frac{3}{2}\rangle
|n\rangle $. In the subspace containing only the $n$-th manifold, the
Hamiltonian takes the form
\begin{widetext}
\begin{equation}\label{h0}
H^{^{0th}}=\left(
\begin{array}{cccc}
\omega n-\frac{9g^{2}}{4\omega } & \frac{\sqrt{3}}{2}\Delta G_{0}(n) & 0 & 0
\\
\frac{\sqrt{3}}{2}\Delta G_{0}(n) & \omega n-\frac{g^{2}}{4\omega } & \Delta
G_{0}(n) & 0 \\
0 & \Delta G_{0}(n) & \omega n-\frac{g^{2}}{4\omega } & \frac{\sqrt{3}}{2}%
\Delta G_{0}(n) \\
0 & 0 & \frac{\sqrt{3}}{2}\Delta G_{0}(n) & \omega n-\frac{9g^{2}}{4\omega }%
\end{array}%
\right).
\end{equation}%
\end{widetext}
The corresponding eigenvalues
and eigenvectors are straightforwardly given by respectively
\begin{eqnarray}  \label{zeroenergy}
\varepsilon _{1,n} &=&\omega n-\frac{5g^{2}}{4\omega }-\frac{1}{2}%
B_{n}-2\chi _{1,n},  \notag  \label{zeroenergy} \\
\varepsilon _{2,n} &=&\omega n-\frac{5g^{2}}{4\omega }+\frac{1}{2}%
B_{n}-2\chi _{2,n},  \notag \\
\varepsilon _{3,n} &=&\omega n-\frac{5g^{2}}{4\omega }-\frac{1}{2}%
B_{n}+2\chi _{1,n},  \notag \\
\varepsilon _{4,n} &=&\omega n-\frac{5g^{2}}{4\omega }+\frac{1}{2}%
B_{n}+2\chi _{2,n},
\end{eqnarray}%
and
\begin{eqnarray}
|\varphi _{1,n}\rangle &\varpropto &\left(
\begin{array}{c}
-1 \\
K_{1,n} \\
-K_{1,n} \\
1%
\end{array}%
\right) ,|\varphi _{2,n}\rangle \varpropto \left(
\begin{array}{c}
1 \\
-K_{2,n} \\
-K_{2,n} \\
1%
\end{array}%
\right) ,  \notag \\
|\varphi _{3,n}\rangle &\varpropto &\left(
\begin{array}{c}
-1 \\
K_{3,n} \\
-K_{3,n} \\
1%
\end{array}%
\right) ,|\varphi _{4,n}\rangle \varpropto \left(
\begin{array}{c}
1 \\
-K_{4,n} \\
-K_{4,n} \\
1%
\end{array}%
\right) ,  \label{eigenstates}
\end{eqnarray}%
where
\begin{equation}
K_{i,n}=\{%
\begin{array}{c}
\frac{1}{\sqrt{3}B_{n}}[-\frac{2g^{2}}{\omega }-(-1)^{i}B_{n}+4\chi
_{i,n}],(i=1,2) \\
\frac{1}{\sqrt{3}B_{n}}[-\frac{2g^{2}}{\omega }-(-1)^{i}B_{n}-4\chi
_{i-2,n}],(i=3,4)%
\end{array}%
,  \label{ki}
\end{equation}%
and
\begin{equation*}
\chi _{i,n}=\sqrt{\frac{g^{4}}{4\omega ^{2}}+(-1)^{i}\frac{g^{2}}{4\omega }%
B_{n}+\frac{B_{n}^{2}}{4}}(i=1,2)
\end{equation*}
with $B_{n}=\Delta G_{0}(n)$. Interestingly, the zeroth-order approximation is similar
to the adiabatic approximation in the two-qubit system~\cite{agarwal}, where the transition
between different manifolds is not considered, and the $n$th state is only limited to the same $n$-th manifold.

The validity of the zeroth-order approximation is restricted to the large detuning regime $\Delta/\omega\ll1$.
In the zero detuning limiting, $\Delta =0$, within the same manifold $n$, $|\pm \frac{3}{%
2}\rangle |n\rangle $ and $|\pm \frac{1}{2}\rangle |n\rangle $ are nearly
degenerate. For a large
detuning $\Delta/\omega\ll1 $, it is reasonable to consider the qubit states with the same $n$
manifold coupled by the interaction.
Especially, for a strong coupling strength $g/\omega\gg\Delta/\omega$, the diagonal
terms of the approximated Hamiltonian in Eq.(~\ref{h0}) play a more dominant role than the off-diagonal terms dependent on $\Delta$.
And the high order terms in Eq.(~\ref{H_11}) still can be neglected even in the strong coupling regimes.
Hence, the zeroth-order approximation is expected to work well from weak to strong coupling regimes for the large detuning case $\Delta/\omega\ll1$.

\begin{center}
\begin{figure*}[tbp]
\includegraphics[scale=0.8]{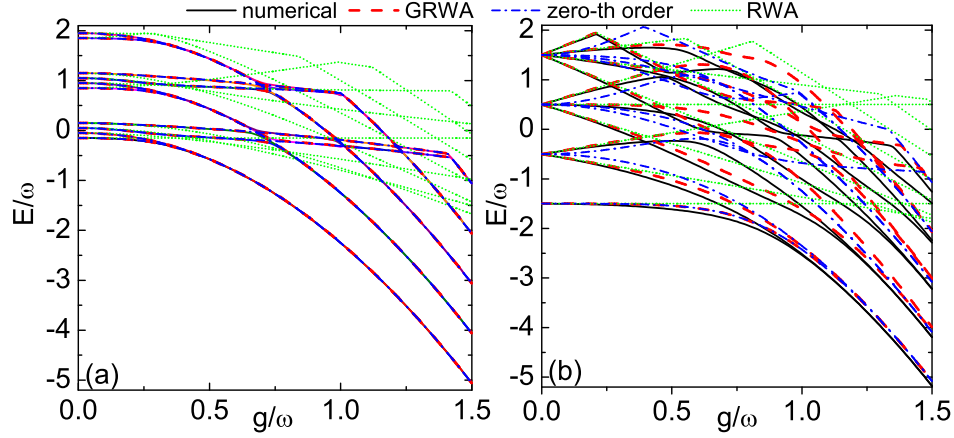}
\caption{(Color online) Energy levels obtained by the GRWA (dashed
lines) for different $\Delta/\protect\omega =0.1$ (a), and $\Delta/\protect%
\omega =1$ (b). The energies by the numerically exact diagonalization (
solid lines), results of RWA (short dotted lines) and results obtained by
the zeroth-order approximation (dashed dotted lines) are plotted for
comparison.}
\label{energy level}
\end{figure*}
\end{center}

The zeroth-order energy spectrum is plotted in Fig.~\ref%
{energy level} with dash-dotted lines.
In large detuning regime $\Delta/\omega =0.1$,
the zeroth-order results agree well with the numerical ones from weak to
strong coupling regimes in Fig.~\ref{energy level}(a). But the
RWA fails to give correct energies as the coupling strength $g/\omega$
increases. Because of the coupling of the qubit and the original oscillator, the latter should be displaced. Thus, the displaced oscillator state in
the zeroth-order approximation, $|n\rangle _{j}=\exp [\frac{jg}{\omega }%
(a^{\dagger}-a)]|n\rangle (j=\pm \frac{3}{2},\pm \frac{1}{2})$, plays a more
important role than the original oscillator state $|n\rangle$ in the RWA,
resulting in more accurate eigen-energies in Eq. (~\ref{zeroenergy}) and
eigenfunctions in Eq. (~\ref{eigenstates}). However, there is a noticeable
deviation of the zeroth-order approximated results for the resonance case $%
\Delta /\omega =1$, indicating that the higher-order terms in Eq. (\ref{H_11}%
) should be taken into account. Physically, qubit states with different $n$ manifolds should be coupled by the interactions.

\textsl{First-order approximation:} Keeping the linear terms in $a$ and $%
a^{\dag }$ and neglecting all higher order terms in the interaction Hamiltonian $H_{1}$(\ref{H_11}) gives
\begin{equation}
H_{1}=\Delta \{J_{x}G_{0}\left( a^{\dagger }a\right) +iJ_{y}[F_{1}\left(
a^{\dagger }a\right) a^{\dagger }-aF_{1}\left( a^{\dagger }a\right) ]\}.
\label{H_1}
\end{equation}%
The term $F_{1}\left( a^{\dagger }a\right) a^{\dagger }$ describes the
photon hopping from state $\left\vert n\right\rangle $ to $\left\vert
n+1\right\rangle $. It is reasonable to set $\left\langle n+1\right\vert
R_{n+1,n}a^{\dagger }\left\vert n\right\rangle =\left\langle n+1\right\vert
F_{1}\left( a^{\dagger }a\right) a^{\dagger }\left\vert n\right\rangle $ by
\begin{eqnarray}
R_{n+1,n} &=&\left\langle n+1\right\vert \sinh \left[ \frac{g}{\omega }%
\left( a^{\dagger }-a\right) \right] \left\vert n\right\rangle /\sqrt{n+1}
\notag \\
&=&\frac{1}{n+1}\frac{g}{\omega }e^{-\frac{g^{2}}{2\omega ^{2}}}L_{n}^{1}(%
\frac{g^{2}}{\omega ^{2}}).  \label{Rn}
\end{eqnarray}%
Similarly, the term $aF_{1}\left( a^{\dagger }a\right) $ only has non-vanishing
element $\langle n\left\vert aF_{1}\left( a^{\dagger }a\right) |n+1\right\rangle $.
It follows that the term $F_{1}\left( a^{\dagger
}a\right) a^{\dagger }$\ creates and $aF_{1}\left( a^{\dagger }a\right) $
eliminates a single photon of the cavity. The physics process is similar to that described in
the RWA model, which facilitates the further analytic treatment.

The Hamiltonian now is $H^{1st}=H_{0}^{^{\prime }}+H_{1}^{^{\prime }}$:
\begin{eqnarray}
H_{0}^{^{\prime }} &=&\omega a^{\dagger }a-\frac{g^{2}}{\omega }%
J_{z}^{2}+\Delta \beta J_{x}, \\
H_{1}^{^{\prime }} &=&\Delta J_{x}[G_{0}\left( a^{\dagger }a\right) -\beta
]+iJ_{y}\Delta \lbrack F_{1}\left( a^{\dagger }a\right) a^{\dagger
}-aF_{1}\left( a^{\dagger }a\right) ],  \notag
\end{eqnarray}%
where $\beta =G_{0}\left( 0\right) =e^{-\frac{g^{2}}{2\omega ^{2}}}$.

Since the qubit and cavity in the noninteracting part $H_{0}^{^{\prime }}$ are
decoupled, we apply a unitary transformation $S$ to diagonalize the qubit
part in $H_{0}^{^{\prime }}$
\begin{equation}
S=\left(
\begin{array}{cccc}
-\frac{1}{C_{1}} & \frac{1}{C_{2}} & -\frac{1}{C_{3}} & \frac{1}{C_{4}} \\
\frac{K_{1}}{C_{1}} & -\frac{K_{2}}{C_{2}} & \frac{K_{3}}{C_{3}} & -\frac{%
K_{4}}{C_{4}} \\
-\frac{K_{1}}{C_{1}} & -\frac{K_{2}}{C_{2}} & -\frac{K_{3}}{C_{3}} & -\frac{%
K_{4}}{C_{4}} \\
\frac{1}{C_{1}} & \frac{1}{C_{2}} & \frac{1}{C_{3}} & \frac{1}{C_{4}}%
\end{array}%
\right) ,
\end{equation}%
where $K_{i}$ has been defined in Eq.(~\ref{ki}) for $n=0$, and the
normalized parameter is $C_{i}=\sqrt{2+2K_{i}^{2}}$.
The corresponding eigenvalues are $\varepsilon_{i,0}$ in Eq.(~\ref{zeroenergy}).

In terms of  the transformation $S^{\dagger}H_{1}^{^{\prime }}S$,
the Hamiltonian $H^{1st}$ of the three-qubit Dicke model can be approximated as
\begin{widetext}
\begin{eqnarray}
H_{\texttt{GRWA}} &=&\omega a^{\dagger }a+\mu _{1}(a^{\dagger }a)|-\frac{3}{2}\rangle \langle -\frac{%
3}{2}|+\mu _{2}(a^{\dagger }a)|-\frac{1}{2}\rangle \langle -\frac{1}{2}|+\mu _{3}(a^{\dagger }a)|\frac{1}{2%
}\rangle \langle \frac{1}{2}|+\mu _{4}(a^{\dagger }a)|\frac{3}{2}\rangle \langle \frac{3}{2}%
|  \nonumber  \label{GRWAham} \\
&&+\Delta F_{1}\left( a^{\dagger }a\right) [\frac{-\sqrt{3}K_{2}+K_{1}(\sqrt{%
3}+2K_{2})}{C_{1}C_{2}}(a|-\frac{1}{2}\rangle \langle -\frac{3}{2}|+h.c)  \nonumber \\
&&+\frac{-\sqrt{3}K_{3}+K_{2}(\sqrt{3}-2K_{3})}{C_{2}C_{3}}(a|\frac{1}{2}%
\rangle \langle -\frac{1}{2}|+h.c)
\nonumber \\
&&+\frac{-\sqrt{3}K_{4}+K_{3}(\sqrt{3}+2K_{4})}{C_{3}C_{4}}(a|\frac{3}{2}%
\rangle \langle \frac{1}{2}|+h.c)],
\end{eqnarray}
\end{widetext}
where $\mu _{i}(a^{\dagger }a)=\varepsilon _{i,0}-\Delta [G_{0}\left( a^{\dagger }a\right) -\beta ]\frac{2K_{i}[\sqrt{3%
}-(-1)^{i}K_{i}]}{_{C_{i}^{2}}}$.
There are only the energy-conserving terms $(a|-\frac{1}{2}\rangle \langle -%
\frac{3}{2}|+h.c)$, $(a|\frac{1}{2}\rangle \langle -\frac{1}{2}|+h.c)$, and $%
(a|\frac{3}{2}\rangle \langle \frac{1}{2}|+h.c)$ with renormalized
coefficients, originating from the CRW terms $iJ_{y}[F_{1}\left(
a^{\dagger }a\right) a^{\dagger }-aF_{1}\left( a^{\dagger }a\right)
]$. The dominated effect of the original CRW terms is considered
here. Because it is the three-qubit Dicke model Hamiltonian in the same
RWA form with renormalized coefficients, the present approach
essentially borrows the basic idea of the GRWA  proposed by Irish
for the one-qubit model~\cite{irish}.

Note that the individual bosonic creation (annihilation) operator $%
a^{\dagger }\left( a\right) $ appears in the GRWA, so the qubits
states with different oscillator number $n$, $n\pm1$ and $n+2$ are coupled with each other. In the basis of $|-\frac{3}{2}\rangle |n+2\rangle $, $|-\frac{1}{2}\rangle
|n+1\rangle $, $|\frac{1}{2}\rangle |n\rangle $ and $|\frac{3}{2}\rangle
|n-1\rangle $ ($n>0$), the Hamiltonian $H_{\texttt{GRWA}}$ can be written in the matrix form as
\begin{widetext}
\begin{equation}
H_{\texttt{GRWA}}=\left(
\begin{array}{cccc}
\omega (n+2)+\mu _{1}(n+2) & \Delta R_{n+1,n+2}^{\prime } & 0 & 0 \\
\Delta R_{n+1,n+2}^{\prime } & \omega (n+1)+\mu _{2}(n+1) & \Delta R_{n,n+1}^{\prime }
& 0 \\
0 & \Delta R_{n,n+1}^{\prime } & \omega n+\mu _{3}(n) & \Delta R_{n-1,n}^{\prime } \\
0 & 0 & \Delta R_{n-1,n}^{\prime } & \omega (n-1)+\mu _{4}(n-1)%
\end{array}%
\right) ,
\end{equation}
\end{widetext}with $R_{n+1,n+2}^{\prime }=\frac{-\sqrt{3}K_{2}+K_{1}(\sqrt{3}%
+2K_{2})}{C_{1}C_{2}}R_{n+1,n+2}\sqrt{n+2}$, $R_{n,n+1}^{\prime }=\frac{-%
\sqrt{3}K_{3}+K_{2}(\sqrt{3}-2K_{3})}{C_{2}C_{3}}R_{n,n+1}\sqrt{n+1}$ and $%
R_{n-1,n}^{\prime }=\frac{-\sqrt{3}K_{4}+K_{3}(\sqrt{3}+2K_{4})}{C_{3}C_{4}}%
R_{n-1,n}\sqrt{n}$.

To this end, the GRWA can be also performed analytically  without more efforts than
those in the original Hamiltonian $H_{\texttt{RWA}}$ in Eq.(\ref{RWA}).
The displaced oscillator states $|n\rangle _{m}$, $|n\pm
1\rangle _{m}$ and $|n+2\rangle _{m}$ depend upon the Dicke state $%
|j,m\rangle $, and are definitely different from both the RWA ones and the zeroth-order approximations where
only the state $|n\rangle_{m}$ is considered. Hence, as $\Delta/\omega$ increases,
the first-order correction provides an efficient, yet accurate analytical solution.

The ground-state energy for the ground state $|-\frac{3}{2}\rangle |0\rangle
$ is
\begin{equation}
E_{0}=-\frac{5g^2}{4\omega}-\frac{\Delta }{2}e^{-\frac{g^{2}}{2\omega ^{2}}}-2\chi _{1,0}.
\end{equation}
The first and second excited energies $\{E_{0}^{k}\}$ ($k=1,2$) can be given
by expanding the GRWA Hamiltonian in the basis $|-\frac{3}{2}\rangle
|1\rangle$ and $|-\frac{1}{2}\rangle |0\rangle$
\begin{equation}
H_{\mathtt{GRWA}}=\left(
\begin{array}{cc}
\omega +\mu _{1}(1) & \Delta R_{0,1}^{\prime } \\
\Delta R_{0,1}^{\prime } & \mu _{2}(0)%
\end{array}%
\right) .
\end{equation}%
Similarly, $H_{\mathtt{GRWA}}$ is given in terms of $|-\frac{3}{2}\rangle
|2\rangle $, $|-\frac{1}{2}\rangle |1\rangle $, $|\frac{1}{2}\rangle
|0\rangle $ as
\begin{equation}
H_{\mathtt{GRWA}}=\left(
\begin{array}{ccc}
2\omega +\mu _{1}(2) & \Delta R_{1,2}^{\prime } & 0 \\
\Delta R_{1,2}^{\prime } & \omega +\mu _{2}(1) & \Delta R_{0,1}^{\prime } \\
0 & \Delta R_{0,1}^{\prime } & \mu _{3}(0)%
\end{array}%
\right) ,
\end{equation}%
which provides three analytical excited energies $\{E_{0}^{k}\}$ ($k=3,4,5$).

Energies obtained by the GRWA are presented in dashed lines in Fig.~\ref%
{energy level}. Especially, for the resonance case $\Delta =\omega $, the GRWA results are much better than the zeroth-order results (blue dotted lines) in Fig.\ref{energy level}(b). It ascribes to the effect of the coupling between states
with different manifolds. Our approach is basically a perturbative expansion
in terms of $\Delta/\omega$. As the increase of the $\Delta /\omega$ ,
the high order terms in Eq.(5) still cannot be neglected in the intermediate and strong coupling regimes.
So the GRWA works reasonably well in the ultra-strong coupling regime $g/\omega<0.3$ at resonance.
Interestingly, the level crossing is present in both the GRWA results
and the exact ones. The RWA requires weak coupling due to the complete neglect of the CRW terms, which are
qualitatively incorrect as the coupling strength increases. So the GRWA includes the dominant contribution of the CRW
terms, exhibiting substantial improvement
of energy levels over the RWA one. The RWA fails in particular to describe
the eigenstates, which should be more sensitive in the quantum entanglement
presented in the next section.

\section{Quantum entanglement}

In the present three-qubit system, we study the GME for the multipartite entanglement and the concurrence for the bipartite
entanglement. A fully separable three-particle state must contain no entanglement.
If the state is not fully separable, then it
contains some entanglement, but it might be still separable with respect to
two-party configurations. For genuine multiparticle entangled states, all
particles are entangled and therefore GME is   very important   among various definition of entanglements.

We review the basic definitions of GME for the three qubits $A$, $B$, and $C$. A separable state is a mixture of product states with respect to a bipartition $A|BC$, that is $\rho_{A|BC}^{sep}=\sum_{j}p_j|\varphi_A^j\rangle\langle\varphi_A^j|\otimes|\varphi_{BC}^j\rangle\langle\varphi_{BC}^j|$,
 where $p_j$ is a coefficient. Similarly, we denote other separable states for the two other bipartitions as $\rho_{B|AC}^{sep}$ and $\rho_{C|AB}^{sep}$. A biseparable state is a mixture of separable states, and combines the separable states $\rho_{A|BC}^{sep}$, $\rho_{B|AC}^{sep}$, and $\rho_{C|AB}^{sep}$ with respect to all possible bipartitions. Any state that is not a biseparable state is called genuinely multipartite entangled.

Recently, a powerful technique has been advanced to characterize multipartite entanglement using positive partial transpose (PPT) mixtures~\cite{peres}. It is well known that a separable state is PPT, implying that its partial transpose is positive semidefinite.
We denote a PPT mixture of a tripartite state as a convex combination of PPT states $\rho_{A|BC}^{PPT}$, $\rho_{B|AC}^{PPT}$ and $\rho_{C|AB}^{PPT}$
with respect to different bipartitions.
The set of PPT mixtures contains the set of
biseparable states. The advantage of using PPT mixtures instead of biseparable states is that the set of PPT mixtures
can be fully characterized by the linear semidefinite programming (SDP)~\cite{boyd},
which is a standard problem of constrained convex optimization theory.

In order to characterize PPT mixtures, a multipartite state which is not a PPT mixture can be detected by a decomposable entanglement witness $W$~\cite{novo}. The witness operator is defined as $W=P_M+Q_M^{T_M}$ for all bipartitions $M|\bar{M}$, where $P_M$, and $Q_M$ are positive semidefinite operators, and $T_M$ is the partial transpose with respect to $M$. This observable $W$ is positive on all PPT mixtures, but has a negative expectation value on at least one entangled state.
To find a fully decomposable witness for a given state $\rho$, the convex optimization technique SDP
becomes important, since it allows us to optimize over all fully decomposable witnesses.
Hence, a state $\rho$ is a PPT mixture only if the optimization problem~\cite{novo},
\begin{equation}  \label{minm}
\textrm{minimize:} {}{} \mathtt{Tr}(W\rho).
\end{equation}
has a positive solution. If the minimum in Eq. (~\ref{minm}) is negative, $\rho$ is
not a PPT mixture and hence is genuinely multipartite entangled.
We denote the absolute value of the above minimization as $E(\rho)$. For solving the SDP we use the programs YALMIP and SDPT3~\cite{yalmip,program}, which are freely available.

Now we discuss the dynamics of the GME for the three-qubit entanglement.
The initial entangled three-qubit state is chosen as the W state with only one excitation
\begin{equation}
|W\rangle =\frac{1}{\sqrt{3}}(|100\rangle +|010\rangle +|001\rangle ),
\label{initial state}
\end{equation}%
which corresponds to the Dicke state $|D_{3}\rangle =|-\frac{1}{2}%
\rangle $. For the Hamiltonian (~\ref{Ham}) with respect to the rotation around the $%
y$ axis by the angle $\pi/2$, the initial Dicke state can be written as
\begin{equation}
|D_{3}\rangle =\frac{1}{\sqrt{8}}(-\sqrt{3}|-\frac{3}{2}\rangle -|-\frac{1}{2%
}\rangle +|\frac{1}{2}\rangle +\sqrt{3}|\frac{3}{2}\rangle),
\label{initial state1}
\end{equation}%
and the initial cavity state is the vacuum state $|0\rangle $. Based on the
eigenstates $\left\{ |\varphi _{k,n}\rangle\right\} $ and eigenvalues $%
\left\{ E_{n}^{k}\right\} $ in the GRWA and the zeroth-order approximation,
the wavefunction evolves from the initial state as $|\phi (t)\rangle
=\sum_{n,k}e^{-iE_{n}^{k}t}|\varphi _{k,n}\rangle \langle \varphi
_{k,n}|D_{3}\rangle $. And the three-qubit reduced state $\rho (t)$ can be given by
tracing out the cavity degrees of freedom
\begin{equation}
\rho (t)=\texttt{Tr}_{\mathtt{cavity}%
}(|\phi (t)\rangle \langle \phi (t)|).
\end{equation}
We then calculate the absolute value of the minimum $E(\rho )$ to detect the GME by solving the minimum in Eq.(~\ref{minm}).

\begin{figure}[tbp]
\includegraphics[scale=0.45]{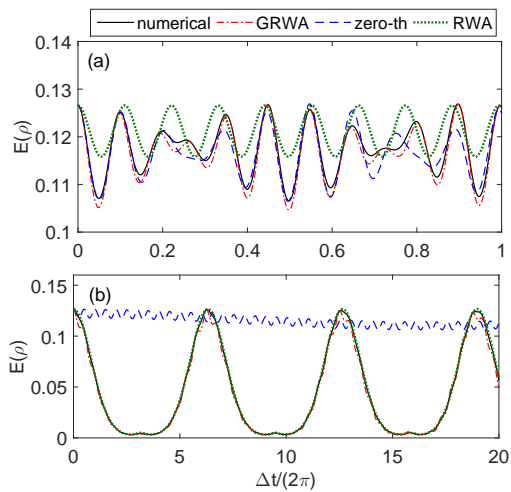}
\caption{(Color online) Dynamics of the GME for three-qubit entanglement
with the initial W state for the ultrastrong-coupling strength $g/\protect%
\omega=0.1$ with the different detuning $\Delta/\protect\omega=0.1$ (a) and $%
\Delta/\protect\omega=1$ (b) by the GRWA method (dash-dotted lines),
numerical method (solid lines), RWA (short-dotted
lines), and the zeroth-order approximation (dashed lines).}
\label{dynamics GEM}
\end{figure}

Fig.~\ref{dynamics GEM} shows the $E(\rho )$ plotted against parameter $\Delta t/(2\pi)$ for different detunings $\Delta/\omega$
for the ultra-strong-coupling strength $g/\omega =0.1$. For comparison, results from numerical exact
diagonalization and RWA are also shown. We observe a quasi-periodic behavior
of the GME dynamics. $E(\rho )$ decays from the initial entangled W state
and falls off to a nonzero minimum value, implying no death of the three-qubit entanglement. The GME dynamics obtained by the
GRWA are consistent with the numerical results, while the RWA results are
qualitatively incorrect for the off-resonance case $\Delta /\omega =0.1$ in
Fig.~\ref{dynamics GEM} (a). The zeroth-order approximation, where only states within
the same manifold are included, works well for the off-resonance case $%
\Delta =0.1$ in Fig.~\ref{dynamics GEM} (a) but not for the on-resonance
case in Fig.~\ref{dynamics GEM} (b). The validity of the GRWA ascribes to the
inclusion of the CRW interaction $iJ_{y}F_{1}\left( a^{\dagger }a\right)
(a^{\dagger }-a)$.

The onset of the decay of the multipartite
entanglement is due to the information loss of qubits dynamics to the cavity.
On the other hand, it is the interaction with the cavity that leads to the
entanglement resurrection. The lost information will be transferred back to the qubit
subsystem after a finite time, which is associated with the ratio between the
coupling strength $g/\omega$ and the level-splitting of qubits $\Delta/\omega$.
As the ratio $g/\Delta$ increases, the contributions of the qubit-cavity interaction become dominant
and the lost entanglement will be transferred quickly from the cavity to qubits with
less revivals time, as shown in Fig~\ref{dynamics GEM} (a).

Moreover, it is significant to study the different behavior of the multipartite entanglement and the bipartite entanglement.
The concurrence characterizes the
entanglement between two qubits. Due to the symmetric Dicke states in the
three-qubit collective model, the concurrence is evaluated in terms of the
expectation values of the collective spin operators as $C=\max
\{0,C_{y},C_{z}\}$, where the quantity $C_{n}$ is defined for a given
direction $n(=y,z)$ as $C_{n}=\frac{1}{2N(N-1)}\{N^{2}-4\langle
S_{n}^{2}\rangle -\sqrt{[N(N-2)+4\langle S_{n}^{2}\rangle
]^{2}-[4(N-1)\langle S_{n}\rangle ]^{2}}\}$~\cite{vidal}. From the dynamical
wavefunction $|\phi (t)\rangle $, we can  easily evaluate the coefficients
for the qubit to remain in the $|j,m\rangle $ state
\begin{equation}\label{zeroprob}
P_{m}^{0th}=\sum_{n=0}^{\infty
}\sum_{k=1}^{4}f_{n}(t)e^{-iE_{n}^{k}t},
\end{equation}%
in  the zeroth-order approximation and
\begin{eqnarray}\label{probability}
P_{m}^{\mathtt{GRWA}} &\approx&\sum_{n}^{\infty
}\sum_{k=1}^{4}f_{n}^{k}(t)(e^{-iE_{n-2}^{k}t}+e^{-iE_{n-1}^{k}t}  \notag \\
&&+e^{-iE_{n}^{k}t}+e^{-iE_{n+1}^{k}t}),
\end{eqnarray}%
in the GRWA. $f_{n}^{k}(t)$ is a dynamical parameter associated with the
initial state and the $k$-th eigenstates for each $n$. From $P_{m}^{\mathtt{GRWA}}$ in Eq.(~\ref{probability}), we
observe energy-level transitions among $E_{n-2}^{k}$, $E_{n\pm 1}^{k}$ and $%
E_{n}^{k}$ in the GRWA, which produce essential improvement of the dynamics
over the zeroth-order ones in Eq.(~\ref{zeroprob}). Since the average value of collective
spin operators can be expressed by $P_m$, such as $4\langle S_{y}^{2}\rangle =4\sqrt{3}({}_{-\frac{3}{2%
}}\langle n-2|n\rangle _{\frac{1}{2}}P_{-\frac{3}{2}}P_{\frac{1}{2}}+{}_{-%
\frac{1}{2}}\langle n-1|n+1\rangle _{\frac{3}{2}}P_{-\frac{1}{2}}P_{\frac{3}{%
2}})-4(P_{-\frac{1}{2}}^{2}+P_{\frac{1}{2}}^{2})+3$, we calculate the concurrence $C$ by the zeroth-order approximation and the GRWA, respectively.

\begin{figure}[tbp]
\includegraphics[scale=0.7]{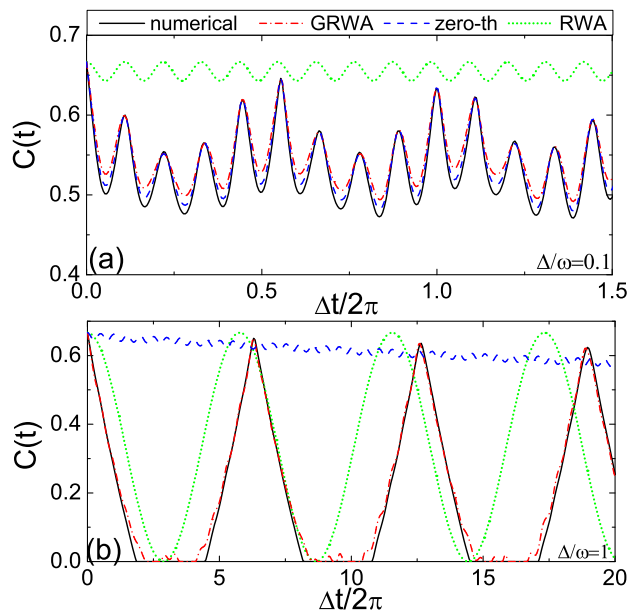}
\caption{(Color online) Dynamics of the concurrence for the qubit-qubit
entanglement with the initial W state for the ultrastrong coupling strength $%
g/\protect\omega=0.1$. The parameters are the same as in Fig.~\protect\ref%
{dynamics GEM}.}
\label{dynamics concurrence}
\end{figure}

We plot the dynamics of the concurrence for different detunings $\Delta /\omega =0.1$ and $1$ in
Fig.~\ref{dynamics concurrence}. The initial W state gives the maximum
pairwise entanglement $C=2/3$ of any Dicke states. Fig.~\ref{dynamics
concurrence} (a) shows that dynamics of the concurrence by the zeroth-order
approximation are similar to the numerical ones in the off-resonance case $%
\Delta /\omega =0.1$, in which the RWA results are invalid. The sudden death
of the bipartite entanglement is observed in the resonance case in Fig.~\ref%
{dynamics concurrence} (b). The dynamics of the concurrence obtained by the
GRWA is similar to the numerical results, exhibiting the disappearance of the
entanglement for a period of time.
However, there is no sudden death of the
entanglement in the RWA case, indicating that the CRW terms are not negligible.

Very interestingly, as shown in Fig.~\ref{dynamics GEM}, the GME  for the three-qubit entanglement never vanishes, in sharp contrast with bipartite entanglement.
During the vanishment of
concurrence, the GME is generally  small but still finite.
It follows  that the  two-qubit state is separable in the system, but the three-qubit state still contains residual entanglement. This may be one
advantage to using GME as a quantum information resource.

\begin{figure}[tbp]
\includegraphics[scale=0.45]{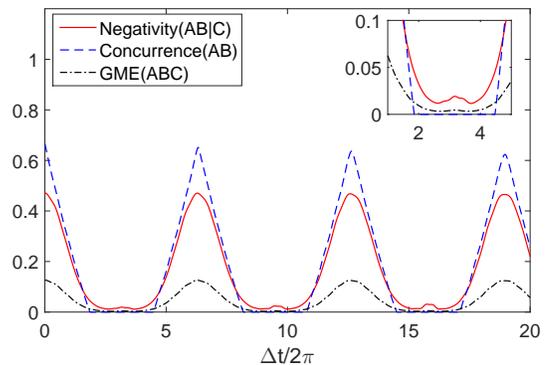}
\caption{(Color online) GME for the three qubits A, B,C (dash-dotted line), negativity for the entanglement with respect to the bipartition $AB|C$ (solid line),
and concurrence between A and B qubits (dashed line) obtained by the numerical method for $%
g/\protect\omega=0.1$ and $\Delta/\protect\omega =1$.}
\label{entanglement}
\end{figure}

Finally, it is significant to clarify why the GME of the tripartite entanglement behaves differently
with the concurrence of the bipartite entanglement. The well-known death of the concurrence
is related to the disappearance of the entanglement in an arbitrary two-qubit subsystem, say A and B, while a deep understanding
is associated with the question of whether there exists entanglement in the three-qubit system. Intuitively, we may think that entanglement is still stored in the bipartition $AB|C$. Negativity is used to detect the entanglement for this bipartition~\cite{vidal2}, which
falls off to a nonzero minimum in Fig.~\ref{entanglement}. It reveals that the state for the bipartition $AB|C$ is not a separable state. Similarly, those states with respect to other bipartitions $AC|B$ and $BC|A$ are not separable. Therefore, the three-qubit state stays in an entangled state and
the GME for the three-qubit entanglement never disappears during the death of the two-qubit entanglement. The theory of the multipartite entanglement is not fully developed and requires more insightful investigations into more- than two-party systems. We highlight here the different features of the multipartite entanglement and bipartite entanglement in the more- than two-qubit system, and have found that the GME is always robust at least in the qubits and single-mode cavity system.

\section{Conclusion}

In this work, we have extended the original GRWA by Irish for the one-qubit Rabi model to the three-qubit Dicke model by the unitary transformation. The zeroth-order approximation, equivalent to the adiabatic approximation,
is suited for arbitrary coupling strengths for the large detuning case. The first-order approximation, also called GRWA,
works well in a wide range of coupling strength even on resonance and much better than the RWA ones. In the GRWA, the effective Hamiltonian with
the CRW interactions is evaluated as the same form of the ordinary RWA one, which facilitates the derivation of the explicit analytic solutions. All eigenvalues and eigenstates can be approximately given.

By the proposed GRWA scheme, we have also calculated the dynamics of concurrence for the bipartite entanglement and the GME for the multipartite entanglement, which are in quantitative agreement with the numerical ones. The well-known sudden death of the two-qubit entanglement is observed by our analytic solution.
An interesting phenomenon of entanglement is that the GME for the three-qubit entanglement decays to the nonzero minimum during the time window in which the two-qubit entanglement disappears, implying that three qubits remain entangled when the two-qubit state is separable.
Our results indicate that the GME is the powerful entanglement to detect quantum correlations in multipartite systems that cannot be described via bipartite entanglement in subsystems of smaller particles.
There still exists many open problems to the theory of entanglement for multipartite systems due to much richer structure of the entanglement in a more- than two-party system. In particular, the dynamical behaviors for two kinds of
entanglement may be explored in the multi-qubit realized in the recent
circuit QED systems in the ultra-strong coupling.

In the end of the preparation of the present work, we noted a recent paper
by Mao et al. ~\cite{mao} for the same model. We should say that the approach
used there is the adiabatic approximation of the present work, i.e., the
zeroth-order approximation.

\section{Acknowledgements}

This work was supported by National Natural Science Foundation of China
(Grants No.11547305, and No.11474256), Chongqing Research Program of Basic Research and
Frontier Technology (Grant No.cstc2015jcyjA00043), and Research Fund for the Central Universities
(Grant No.106112016CDJXY300005).

$^{*}$ Email:yuyuzh@cqu.edu.cn

$^{\dagger}$ Email:qhchen@zju.edu.cn

\end{document}